\documentclass[a4paper,fleqn,usenatbib]{mnras}

\usepackage{newtxtext,newtxmath}

\usepackage[T1]{fontenc}
\usepackage{ae,aecompl}

\usepackage{graphicx} 
\usepackage{mathrsfs} 
\usepackage{natbib}
\usepackage{amsmath}
\usepackage{float}
\usepackage{amssymb}
\usepackage{color}
\usepackage{captcont}
\usepackage{url}
\usepackage{rotating}
\usepackage{longtable,array}
\usepackage{afterpage}
\usepackage{pdflscape}
\usepackage{threeparttablex}
\usepackage{booktabs} 
\usepackage{captcont}

\title{Identification of Young Stellar Variables with KELT for \emph{K2} II: The Upper Scorpius Association}

\author[Ansdell et al.]{Megan Ansdell$^{1}$, 
Ryan J. Oelkers$^{2}$, 
Joseph E. Rodriguez$^{3}$, 
Eric Gaidos$^{4}$,
\newauthor 
Garrett Somers$^{2}$, 
Eric Mamajek$^{5,6}$,
Phillip A. Cargile$^{3}$,
Keivan G. Stassun$^{2,7}$, 
\newauthor  
Joshua Pepper$^{8}$, 
Daniel J. Stevens$^{9}$,
Thomas G. Beatty$^{10,11}$, 
Robert J. Siverd$^{12}$, 
\newauthor 
Michael B. Lund$^{2}$, 
Rudolf B. Kuhn$^{13}$, 
David James$^{14}$, 
B. Scott Gaudi$^{8}$
\\
$^1$Institute for Astronomy, University of Hawai`i at M\={a}noa, 2680 Woodlawn Dr, Honolulu, HI 96822, USA \\
$^2$Department of Physics and Astronomy, Vanderbilt University, 6301 Stevenson Center, Nashville, TN 37235, USA \\
$^3$Harvard-Smithsonian Center for Astrophysics, 60 Garden St, Cambridge, MA 02138, USA \\
$^4$Department of Geology \& Geophysics, University of Hawai`i at M\={a}noa, Honolulu, HI 96822, USA \\
$^5$Jet Propulsion Laboratory, California Institute of Technology, M/S 321-100, 4800 Oak Grove Dr, Pasadena, CA 91109, USA \\
$^6$Department of Physics and Astronomy, University of Rochester, Rochester, NY 14627, USA \\
$^7$Department of Physics, Fisk University, 1000 17th Avenue North, Nashville, TN 37208, USA \\
$^8$Department of Physics, Lehigh University, 16 Memorial Drive East, Bethlehem, PA 18015, USA \\
$^9$Department of Astronomy, The Ohio State University, Columbus, OH 43210, USA \\
$^{10}$Department of Astronomy \& Astrophysics, The Pennsylvania State University, 525 Davey Lab, University Park, PA 16802, USA \\
$^{11}$Center for Exoplanets and Habitable Worlds, The Pennsylvania State University, 525 Davey Lab, University Park, PA 16802, USA \\
$^{12}$Las Cumbres Observatory Global Telescope Network, 6740 Cortona Dr., Suite 102, Santa Barbara, CA 93117, USA \\
$^{13}$South African Astronomical Observatory, PO Box 9, Observatory 7935, South Africa \\
$^{14}$Astronomy Department, University of Washington, Box 351580, Seattle, WA 98195, USA \\
}

\date{Submmitted to MNRAS on 23 July 2017}

\pubyear{2017}

\begin{document}
\label{firstpage}
\pagerange{\pageref{firstpage}--\pageref{lastpage}}
\maketitle

\begin{abstract}
High-precision photometry from space-based missions such as {\it K2} and {\it TESS} enables detailed studies of young star variability. However, because space-based observing campaigns are often short (e.g., 80 days for {\it K2}), complementary long-baseline photometric surveys are critical for obtaining a complete understanding of young star variability, which can change on timescales of minutes to years. We therefore present and analyze light curves of members of the Upper Scorpius association made over 5.5 years by the ground-based Kilodegree Extremely Little Telescope (KELT), which complement the high-precision observations of this region taken by {\it K2} during its Campaigns~2 and 15. We show that KELT data accurately identify the periodic signals found with high-precision {\it K2} photometry, demonstrating the power of ground-based surveys in deriving stellar rotation periods of young stars. We also use KELT data to identify sources exhibiting variability that is likely related to circumstellar material and/or stellar activity cycles; these signatures are often unseen in the short-term {\it K2} data, illustrating the importance of long-term monitoring surveys for studying the full range of young star variability. We provide the KELT light curves as electronic tables in an ongoing effort to establish legacy time-series datasets for young stellar clusters.
\end{abstract}




\section{Introduction}

Young stars have long been known to exhibit diverse photometric variability attributed to a range of physical mechanisms \cite[see an early review in][]{Bertout:1989}. Classical T Tauri stars (CTTS) are characterized by their full disks and ongoing accretion, which translate into an array of complicated light curve features \cite[e.g.,][]{Cody:2014}. In particular, very deep dimming events lasting days to months have been connected to various forms of occulting disk structures \cite[e.g.,][]{Bouvier:1999,Dullemond:2003,Cody:2014,Ansdell:2016a}. Flaring events related to accretion bursts are also common, especially for magnetically active, rapidly rotating stars. More evolved weak-line T Tauri stars (WTTS), which no longer host significant disks or show clear accretion signatures, typically exhibit sinusoidal variability in their light curves, attributable to spots on the stellar surface rotating with the star. These sinusoidal variations can be used to estimate stellar rotation periods, which lengthen over the first few hundred Myr of evolution as stars spin down via magnetized stellar winds that remove angular momentum \citep{Herbst:2007}. Recently, WTTS have also been found to display long-term variability in their brightness, possibly due to magnetic activity cycles \citep{Rodriguez:2017a}.

Thus studying the photometric variability of young stars can give insight into stellar evolution as well as the structure and dynamics of the surrounding circumstellar material as it presumably forms planets. Much of what we know about the variability of young stars comes from all-sky ($\gtrsim$70\%), long-baseline ($\sim$years), high-cadence (10s of minutes) photometric monitoring surveys from the ground (e.g., ASAS, CRTS, WASP, KELT, ASAS-SN; \citealt{Pojamanski:1997,Drake:2009,2006Pollacco,Pepper:2007,Pepper:2012}) as well as generally shorter-term but much higher-precision monitoring by space-based missions (e.g., CoRoT, {\it Spitzer}/IRAC, {\it Kepler}, {\it K2}; \citealt{2004Fazio,2006Baglin,2010Borucki,Howell:2014}).

The Upper Scorpius association (Upper Sco) is a particularly interesting target for studies of young star variability, as it is the nearest star-forming region \cite[145~pc;][]{1999deZeeuw} with a median age \cite[$\approx5$--11~Myr;][]{2002Preibisch,2012Pecaut} that is comparable to the disk dispersal timescale \citep{2011Williams}. Moreover, its population of several hundred stars and brown dwarfs has been well studied at various wavelengths. In particular, its disk population has been characterized in the infrared by {\it Spitzer} \citep{Carpenter:2006} and WISE \citep{Luhman:2012} as well as in the sub-mm by ALMA \citep{2014Carpenter,Barenfeld2016}. Stars in this association have also been surveyed for multiplicity using ground-based adaptive optics \cite[e.g.,][]{2008Kraus,2014Lafren}. Additionally, the re-purposed space-based {\it Kepler} mission, {\it K2} \citep{Howell:2014}, has provided high-precision time-series photometry during its Campaign~2 (C2), leading to numerous studies of young star variability in Upper Sco \cite[e.g.,][]{2015Ripepi,Ansdell:2016a,Scaringi:2016,2017Cody,2017Stauffer}. {\it K2} will visit another field that partially overlaps with Upper Sco in its upcoming Campaign~15 (C15) planned for 23 August to 20 November 2017, providing yet another opportunity to study young star variability in this association.

The space-based {\it K2} observing campaigns last 80 days and the data are not available until months after they are taken. Thus, from {\it K2} data alone, it is challenging to discern variability on timescales much longer than a month. Moreover, because the photometric variability of the targeted stars is often not known in advance, it is difficult to complement {\it K2} observations with simultaneous spectroscopy or multi-passband photometry, which can be critical to identifying the mechanisms driving young star variability \cite[e.g.,][]{Zhang:2015}. Fortunately, long-term ground-based photometric surveys can identify variables in advance of space-based monitoring campaigns as well as extend the observing baselines to nearly a decade or more in many cases.

To this end, we are providing ground-based photometry of young stellar clusters taken by the Kilodegree Extremely Little Telescope (KELT) in an effort to establish legacy time-series datasets for studying young star variability. KELT provides long-baseline ($\lesssim$10 yr) high-cadence (10-20 min) photometry at $\sim$1\% precision for bright ($7<V<11$) stars across the sky, making it well-suited to studying photometric variability of young star-disk systems on timescales of days to years. In Part {\sc I} of this series, we provided KELT light curves of Taurus-Auriga members targeted by {\it K2} \citep{Rodriguez:2017b}; here in Part {\sc II}, we extend this work to Upper Sco members also targeted by {\it K2}. In Section~\ref{sec:selection} we present our sample and in Section~\ref{sec:kelt} we describe the KELT observations. The variability and periodicity of the sample are assessed in Section~\ref{sec:varper}, then discussed in terms of young star-disk variability in Sections~\ref{sec:exotic} and \ref{sec:discussion}. We summarize our work in Section~\ref{sec:sum}.

\section{Data \& Methods}

\subsection{Target Selection}
\label{sec:selection} 

Our sample of candidate Upper Sco members is compiled by merging the catalogues of \cite{Luhman:2012}, \cite{Rizzuto:2015}, and \cite{2016Pecaut}. We also include the stars identified by \citet{Preibisch:1998} as candidate young cluster members, but caution that their Upper Sco membership is highly uncertain due to their weak lithium absorption. These catalogues were selected using proper motions cuts and/or diagnostics of youth (e.g., infrared excess, lithium absorption, X-ray emission). We do not make any additional selection cuts in order to provide as broad a sample as possible.

We find that 148 candidate Upper Sco members have matches in the KELT survey and also fall on working {\it K2} detectors in C2 or C15. Due to the brightness limit of the KELT survey (see Section~\ref{sec:kelt}), these stars are biased toward Upper Sco members with $M_{\star} \gtrsim 0.5~M_{\odot}$. All of these sources have near-infrared counterparts in the Two Micron All-Sky Survey \cite[2MASS;][]{2006Skrutskie}, thus we use 2MASS names for identification throughout this work. To respect the saturation limit in the KELT-South data, we also make an optical magnitude cut of $V>8$~mag, resulting in a final sample of 131 targets presented in Table~\ref{tab:CatalogueInfo}.

\subsection{KELT Observations}
\label{sec:kelt} 

The KELT survey is designed to discover giant planets transiting bright ($7<V<11$~mag) host stars. KELT consists of two telescopes, each with a 42-mm telephoto lens, with KELT-North at the Winer Observatory in Arizona (United States) and KELT-South at the South African Astronomical Observatory in Sutherland (South Africa). Each KELT telescope uses a Paramount ME German equatorial mount that requires 180-degree rotations when passing the meridian. Because the telescope optics are not axisymmetric, the stellar point-spread function (PSF) changes from east to west observing orientations; east and west observations are therefore treated as two separate telescopes during the data reduction process. Together, the two telescopes observe over 70\% of the sky with 10- to 30-min cadence. Each telescope uses a broad $R$-band filter ($\lambda_{\rm eff} = 691$~nm, ${\rm W}_{\rm eff} = 318$~nm; \citealt{Pepper:2007}) and has a $26^\circ\times26^\circ$ field of view with a 23$\arcsec$ pixel scale. The typical photometric error for bright ($7<V<11$~mag) stars is 1\%, however the telescopes also obtain lower-precision photometry of stars as faint as $V \approx 14$~mag \citep{Pepper:2007,Pepper:2012}. See \cite{Siverd:2012} and \cite{Kuhn:2016} for detailed descriptions of the KELT observing strategy and data reduction process.

The KELT data used in this work are from the KELT-South field 26 (KS-26), which is centered at $\alpha =$ 15$^{h}$ 19$^{m}$ 48$^{s}$, $\delta =$ -20$\degr$ 00$\arcmin$ 00$\arcsec$ (J2000). The KELT light curves for all 131 stars in our sample are available as electronic tables via \texttt{Filtergraph} \citep{2013Burger}.\footnote{\url{https://filtergraph.com/kelt_k2} (this link also includes KELT data for other papers in this series; \citealt{Rodriguez:2017b}).} The single-point photometric error on the data depends on the target brightness, but is typically $<0.05$~mag. The first KELT-South season has a known systematic associated with telescope focusing, thus any long-term changes in brightness that are seen only in this first season should be viewed with extreme caution. Due to the large pixel scale of KELT, some of the targets in our sample may also be blended into a single KELT light curve; we flag 27 sources as possible blends in Table~\ref{tab:CatalogueInfo} due to another star with similar brightness ($\pm$1.5~mag) being located within 2\arcmin{} ($\approx5$~KELT pixels).

\subsection{Variability \& Periodicity Tests}
\label{sec:varper}

\begin{figure*}
\begin{center}
\includegraphics[width=10.2cm]{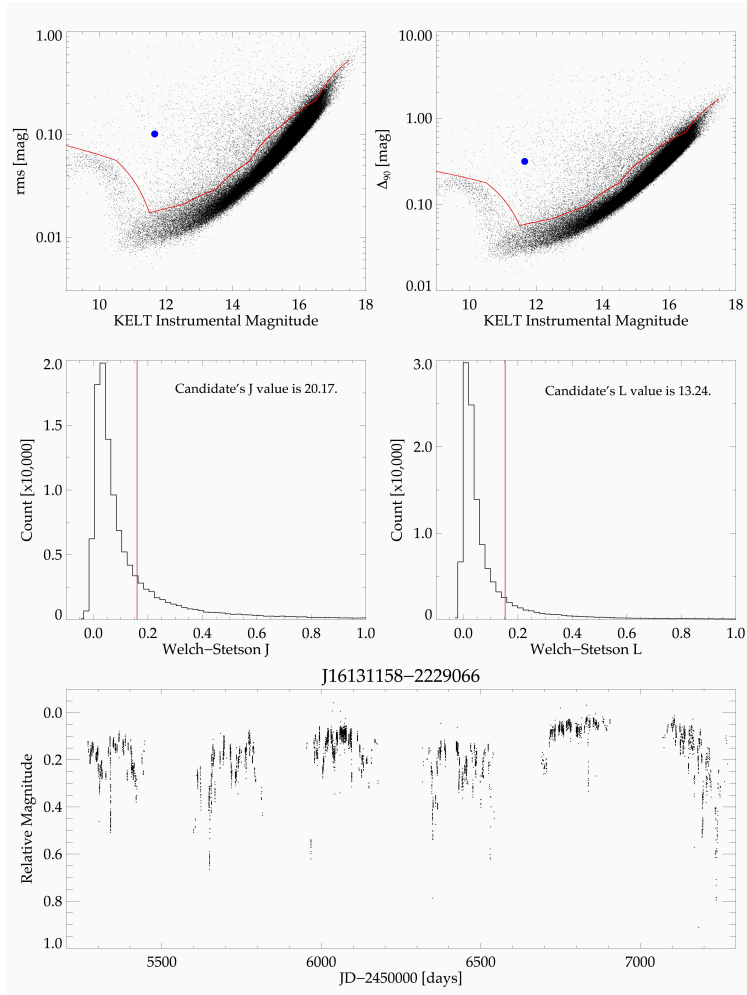}
\caption{Distributions of the four metrics used to identify variable objects in our sample (see Section~\ref{sec:variability}). Top panels: \textit{rms} and $\Delta_{90}$ statistics with their respective $+2\sigma$ cuts (red curves) used to identify candidate variables. Middle panels: Welch-Stetson \textit{J} \& \textit{L} statistics with their respective $+3\sigma$ and $+5\sigma$ cuts (red lines) used to identify candidate variables. Bottom panel: the light curve of an example variable star, J16131158-2229066, whose computed metrics are given in the top (blue points) and middle panels.}
\label{fig:var}
\end{center}
\end{figure*}

\subsubsection{Variability Testing}
\label{sec:variability}

We identify variable sources in our sample by applying four metrics to all the light curves from KS-26, following the work of \cite{Wang:2013} and \cite{Oelkers:2015}. As shown in Figure~\ref{fig:var}, these metrics identify sources with variability amplitudes that are large relative to other stars within the same field, where the variability is not necessarily periodic. We outline each variability metric below, but direct the reader to the previously cited works for more detailed discussions.

We identify candidate variables with light curves exhibiting unusually high dispersion for their magnitude using the \textit{rms} and $\Delta_{90}$ statistics. The \textit{rms} statistic identifies the magnitude range for 68\% of the data points in a light curve, whereas the $\Delta_{90}$ statistic identifies the magnitude range for $90\%$ of the data points in a light curve. Using all the light curves from KS-26, we compute the upper 2$\sigma$ envelopes of both statistics as a function of magnitude, then flag objects in our sample lying above these limits as candidate variables (see top panels in Figure~\ref{fig:var}). In order to define the envelopes using stars with minimal variability, we apply a 2.5$\sigma$ iterative clipping to the sample prior to defining the envelope.

We also compute the Welch-Stetson \textit{J} and \textit{L} statistics \citep{Stetson:1996}, which were designed to identify objects that show correlations in their residuals from the mean with time. As shown in Figure~\ref{fig:var} (middle panels), these statistics are expected to produce a distribution of values centered at or near zero with a one-sided tail containing the variable objects. These statistics perform well on data from the KELT survey due to its high cadence over long time baselines. We compute the $+3\sigma$ cutoff of the tail in \textit{J}, and the $+5\sigma$ cutoff of the tail in \textit{L}, to identify candidate variables. We apply a more stringent cutoff in \textit{L} because this helps to remove the spurious variables that pass the other three metrics but show features of known detector systematics, while still retaining objects that show variations consistent with astrophysical phenomena. In order to calculate the cutoffs for \textit{J} and \textit{L} based on stars showing minimal light curve variations, we remove objects with $J,L > 3$ and perform a $2.5\sigma$ iterative clipping prior to calculating the mean and standard deviation values of the \textit{J} and \textit{L} distributions.

Figure~\ref{fig:var} shows the distributions of the four metrics for all sources in KS-26 as well as the cutoffs for identifying candidate variables in our sample. Sources passing all four metrics are flagged as variables in Table~\ref{tab:CatalogueInfo}, which also gives the rms statistic for each source as an estimate of the variability amplitude. We note that the proximity of KS-26 to the galactic bulge means that these amplitudes may be diluted due to blending.

\subsubsection{Periodicity Testing}
\label{sec:period}

We search each KELT light curve for periodic signals between 0.1 and 150~days using the Lomb-Scargle periodogram \cite[LSP;][]{Lomb:1976, Scargle:1982} as implemented in the \texttt{AstroPy} Python library.\footnote{\url{http://www.astropy.org/}} We reject periods if the power in the periodogram is $<0.02$, or if the period is within 5\% of a solar day or 2\% of its first 10 harmonics ($f$, $f/2$, $f/3$...) to avoid aliases associated with the diurnal cycle. We also reject periods between 26 and 30 days to avoid aliases associated with the lunar cycle, which arise because Moon glow can produce time-correlated noise by regularly increasing the sky background, even for sources separated from the Moon by dozens of degrees. Additionally, we reject periods for which 2$\times$ or 0.5$\times$ the period fails any of the alias checks described above. 

\begin{figure*}
\begin{center}
\includegraphics[width=16.25cm]{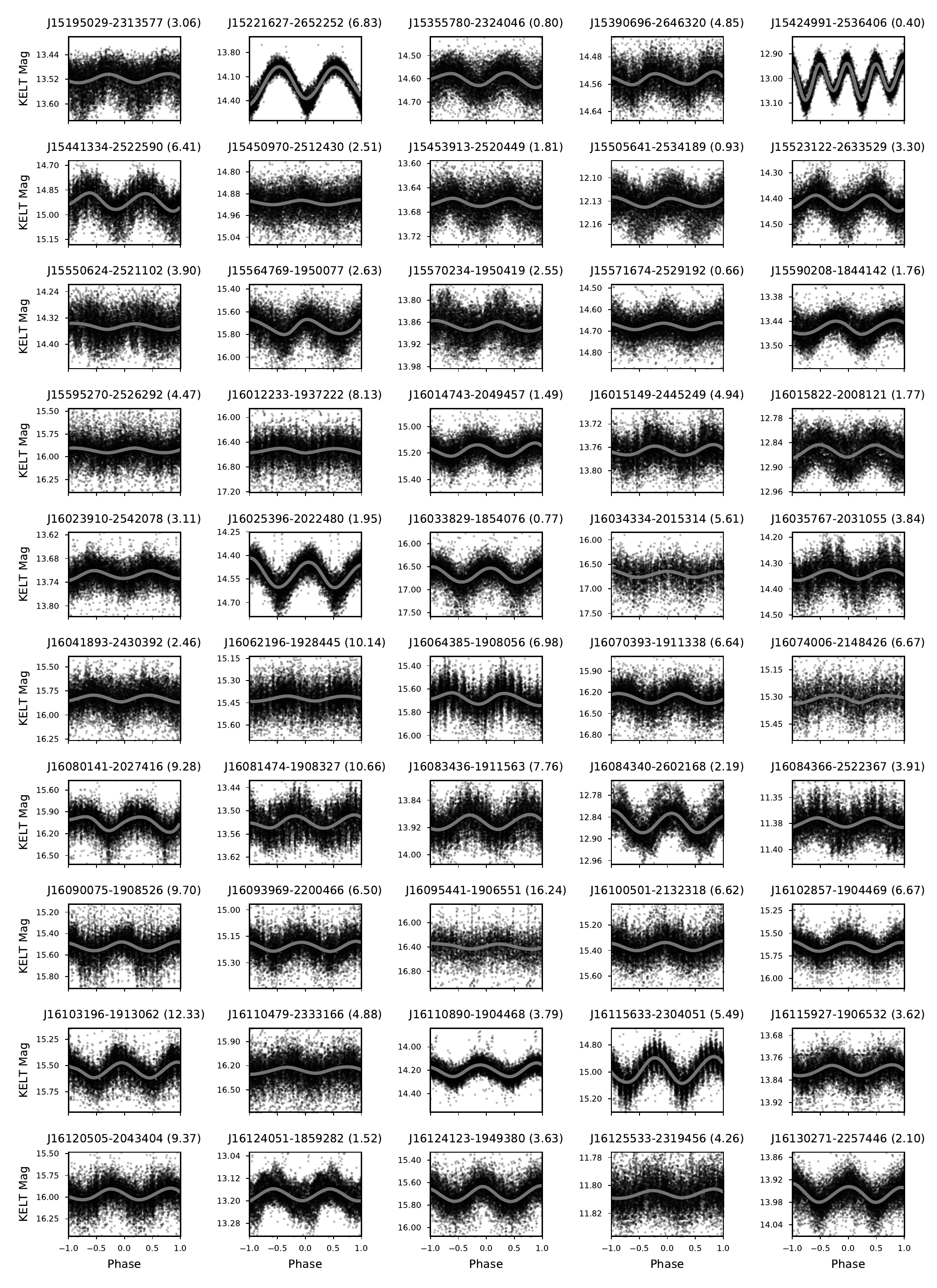}
\caption{Phase-folded KELT light curves (replicated over two full phases) for the sources in our sample exhibiting periodic signals (see Section~\ref{sec:period}). Smoothed medians are shown by the gray lines and the source 2MASS names are given at the top of each panel along with the derived rotation periods (in days) in parentheses. We do not show sources with potentially blended KELT light curves (flagged in Table~\ref{tab:CatalogueInfo}) or the more exotic variables described in Section~\ref{sec:exotic} (see instead Figures~\ref{fig:dip} and \ref{fig:LT}).}
\label{fig:per}
\end{center}
\end{figure*}

The period with the highest power fulfilling the above criteria is listed in Table~\ref{tab:CatalogueInfo} for each target. If no period is listed, then no significant period could be found for that target. The phase-folded light curves for the sources with identified periods are shown in Figure~\ref{fig:per}. We do not report any lower-power periods fulfilling the above criteria, as we find that the highest-power periods derived from KELT data consistently match the periods derived from the much higher-precision {\it K2}/C2 data (see Section~\ref{sec:k2}). As an additional check, we execute a boot-strap analysis using 1000 Monte-Carlo iterations, where for each iteration the observation dates are unchanged but the magnitude values are randomly shuffled \citep{2012Henderson}. We calculate the LSP for each iteration and record the peak power; if the maximum peak power is larger than the power of the period found using the original light curve, we reject the period as a false positive signal. We note that none of the periods derived above were rejected during this additional screening process.

We find periodic signals in the KELT data for 89 sources in our sample. The vast majority of these periodic signals are consistent with the rotation of spots on the surfaces of low-mass stars that are $\approx10$~Myr old. More specifically, they have periods of $\approx0.5$--10 days with amplitudes of a few percent \cite[e.g.,][]{Herbst:2007}. The handful of targets exhibiting more exotic variability are discussed in the following section.

\renewcommand{\thefigure}{\arabic{figure} (Cont.)}
\addtocounter{figure}{-1}

\begin{figure*}
\begin{center}
\includegraphics[width=16cm]{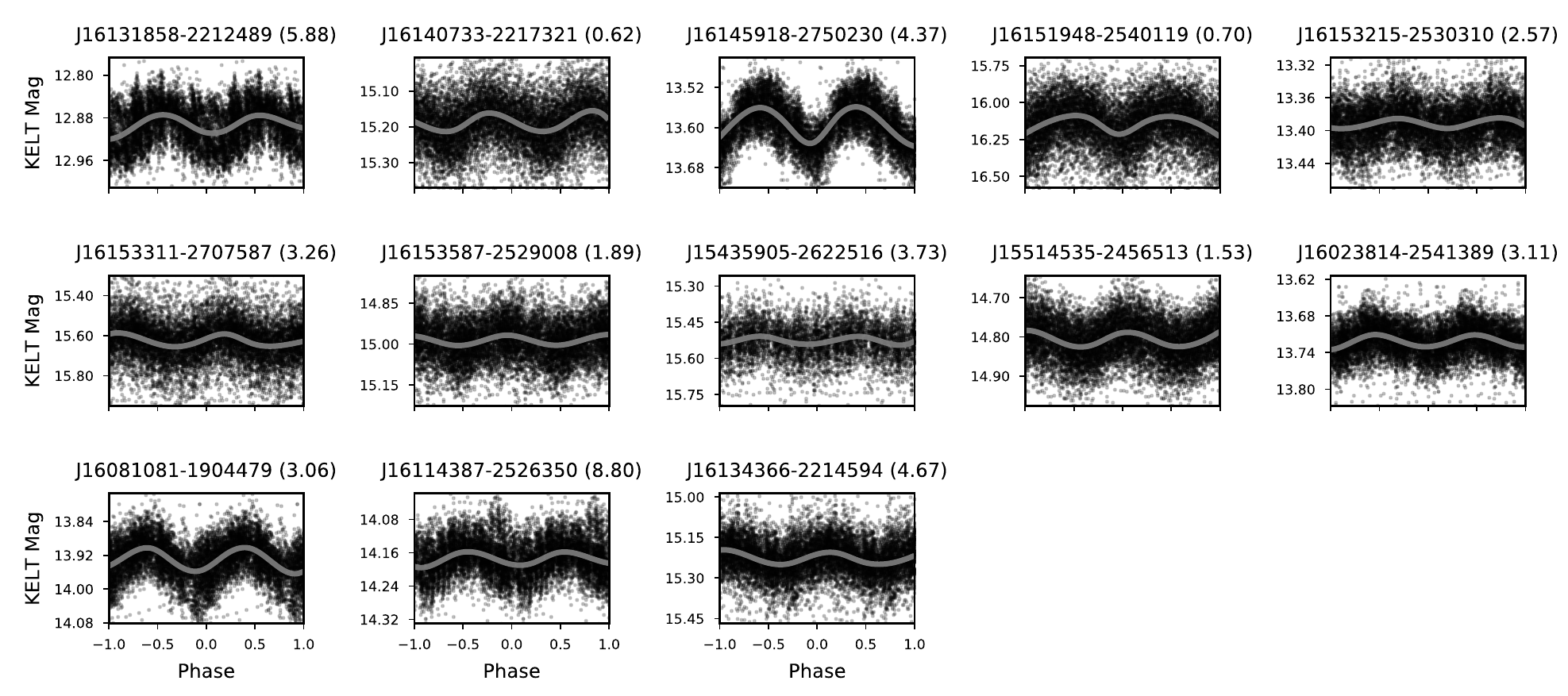}
\caption{}
\end{center}
\end{figure*}

\renewcommand{\thefigure}{\arabic{figure}}

\section{Interesting Variables}
\label{sec:exotic}

\subsection{Dippers \& UXORs}
\label{sec:dippers}

\begin{figure*}
\begin{centering}
\includegraphics[width=15cm]{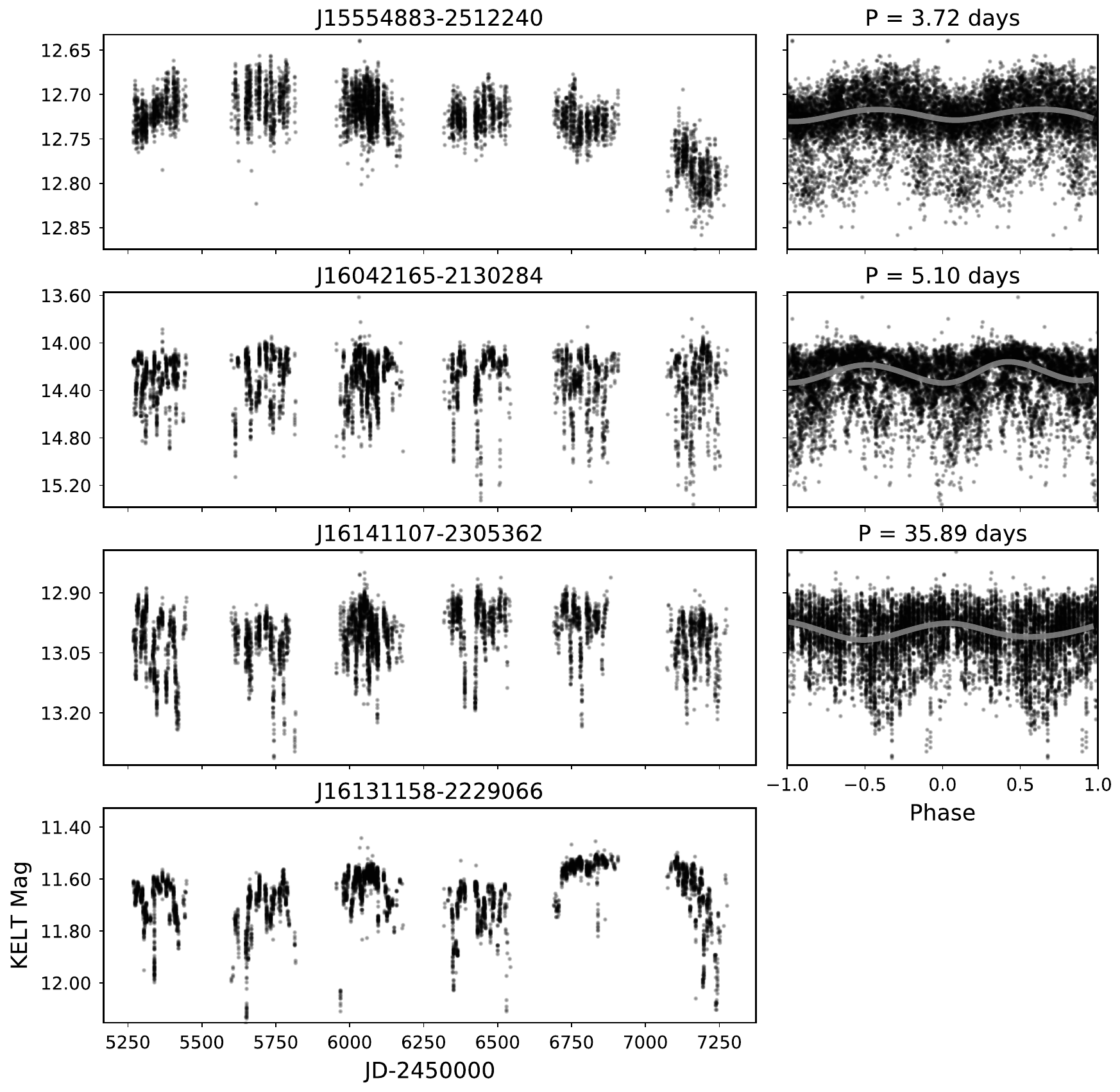}
\caption{The full KELT light curves (left) for the four dipper/UXOR candidates discussed in Section~\ref{sec:dippers}; source 2MASS names are given at the top of each panel. Phased light curves (right; replicated over two full phases) are provided for the three sources that also show periodic variability; smoothed medians are shown by the gray lines and derived rotation periods are given at the top of each panel.}
\label{fig:dip}
\end{centering}
\end{figure*}

Four sources in our sample exhibit ``dipper" or ``UXOR" behavior in their KELT light curves, as shown in Figure~\ref{fig:dip}. Dippers are identified by near-constant maximum brightness levels, punctuated by dimming events that are typically $\sim$0.1--1.0~mag in depth and last for a couple of days. Dippers are thought to result from occultations of T Tauri stars by circumstellar material orbiting in the inner disk \cite[e.g.,][]{Bouvier:1999,Cody:2014,McGinnis:2015,Ansdell:2016a,Bodman:2016}. UXOR variables are typically disk-hosting Herbig Ae stars whose light curves also exhibit dimming events, though the events tend to be deeper ($\sim$1--2~mag) and last longer \cite[i.e., weeks to months;][]{Waters:1998,Dullemond:2003} compared to dippers. It is unclear whether dipper and UXOR variability are produced by the same physical mechanism(s), although both classes of objects are thought to be related to transiting circumstellar dust. Indeed, all four of the dipper/UXOR candidates host disks (based on WISE excess), but interestingly cover a range of disk types \cite[full, transitional, debris;][]{Luhman:2012}. Three of these variables (J15554883-2512240, J16131158-2229066, J16141107-2305362) are newly identified with KELT data, while one dipper (J16042165-2130284) was previously identified in {\it K2}/C2 data \citep{Ansdell:2016b}. Each source is discussed below.

{\bf J15554883-2512240} exhibits aperiodic dimming events $\approx0.1$~mag in depth, superposed on a periodic signal of 3.72~days, in all six KELT-South seasons. The star also maintains a near-constant maximum brightness level of $\approx12.7$~mag until the last KELT season, during which it steadily dims by $\approx0.1$~mag. Interestingly, the source does not exhibit dipper behavior in its {\it K2}/C2 light curve, although dippers are known to change their behavior over time \cite[e.g.,][]{McGinnis:2015,Ansdell:2016a}. The star has a G3V spectral type and hosts a ``debris/evolved transitional" disk \citep{Luhman:2012}. This disk was not detected in a sub-mm ALMA survey of Upper Sco \citep{Barenfeld2016} and the disk inclination remains unknown.

{\bf J16042165-2130284} and J16042097-2130415 are blended in both their KELT and {\it K2}/C2 images, with their combined light curves exhibiting dipper behavior in both datasets. However, because the sources have a sufficiently large projected separation ($\approx16$\arcsec), \cite{Ansdell:2016b} were able to manually extract their individual {\it K2}/C2 light curves from the {\it Kepler} Target Pixel File using custom photometric apertures. They found that J16042165-2130284 is the source of the dipper behavior, with J16042097-2130415 contributing little to the photometric variability. Indeed, J16042097-2130415 does not appear to host a disk based on its lack of infrared excess in WISE pass-bands \citep{Luhman:2012}. In contrast, J16042165-2130284 has a K2V spectral type and hosts a well-characterized face-on transition disk \citep{Mathews2012,Zhang2014}. The system, which is also known simply as J1604, has a KELT light curve that shows a constant maximum brightness level of $\approx14.0$~mag, punctuated by aperiodic dimming events $\approx1$~mag in depth, and superposed on a 5.10-day periodic signal, consistent with the variability seen in its {\it K2}/C2 light curve \citep{Ansdell:2016b}. The KELT data confirm that this target has been exhibiting dipper-like behavior for at least the last 5.5~years.

{\bf J16141107-2305362} exhibits a constant maximum brightness level of $\approx12.9$~mag in its KELT light curve with dimming events $\approx0.2$~mag in depth. The star has a K2V spectral type and hosts a full disk \citep{Luhman:2012}, which has also been detected in the sub-mm with ALMA  \citep{2014Carpenter}. Its {\it K2}/C2 light curve shows variable behavior common to young star-disk systems, but is dominated by a 2.8-day periodic signal that is likely the rotation period of the star. The KELT data, however, reveal a longer periodic signal at 35.89~days, which may be related to a stellar activity cycle. This target also has a close stellar companion with a projected separation of 32~AU \citep{2009Metchev}.
 
{\bf J16131158-2229066} has a KELT light curve that exhibits aperiodic dimming events $\approx0.4$~mag in depth, super-imposed on a varying maximum brightness level with no detectable periodic signal. The star has an A8III/IV spectral type, hosts a full disk, and exhibits clear signatures of ongoing accretion \citep{Luhman:2012}. These features make J16131158-2229066 more consistent with UXOR variables rather than dippers. Its inner disk also has an inclination of $\approx60^{\circ}$ \citep{2017Lazareff}, which may provide the obscuring circumstellar material causing the dimming events. Its {\it K2}/C2 light curve shows similar variability, but this star was not identified in previous dipper/UXOR searches due to concerns with saturation \citep{Ansdell:2016a}. \cite{2015Ripepi} also found high-frequency (30-40~d$^{-1}$), low-amplitude (ppt) variability consistent with $\delta$~Scuti pulsations using {\it K2}/C2 short-cadence data. These pulsation signatures are not seen in the KELT data due to the insufficient cadence and sensitivity of the survey.

Three sources in our sample (J16035767-2031055, J16090075-1908526, J16100501-2132318) were previously identified as dippers from their {\it K2}/C2 light curves \citep{Ansdell:2016a}, but are not identified as such by our KELT data. This is likely due to a combination of their optical faintness and moderate dimming, rather than a change in their dipping behavior. J16035767-2031055 is faint ($V=12.8$) with relatively shallow (0.1~mag) dips in its {\it K2}/C2 light curve. J16090075-1908526 shows deeper (0.2~mag) dips, but is likely too faint ($V=13.8$) for KELT to clearly detect this level of variability (although its KELT light curve does show hints of dipper behavior; see Figure~\ref{fig:per}). Similarly, J16100501-2132318 has extremely shallow (0.03~mag) dips in its {\it K2}/C2 light curve and is also faint ($V=13$~mag), making its variability difficult to detect with KELT.

\afterpage{
\begin{figure*}
\begin{centering}
\includegraphics[width=15.5cm]{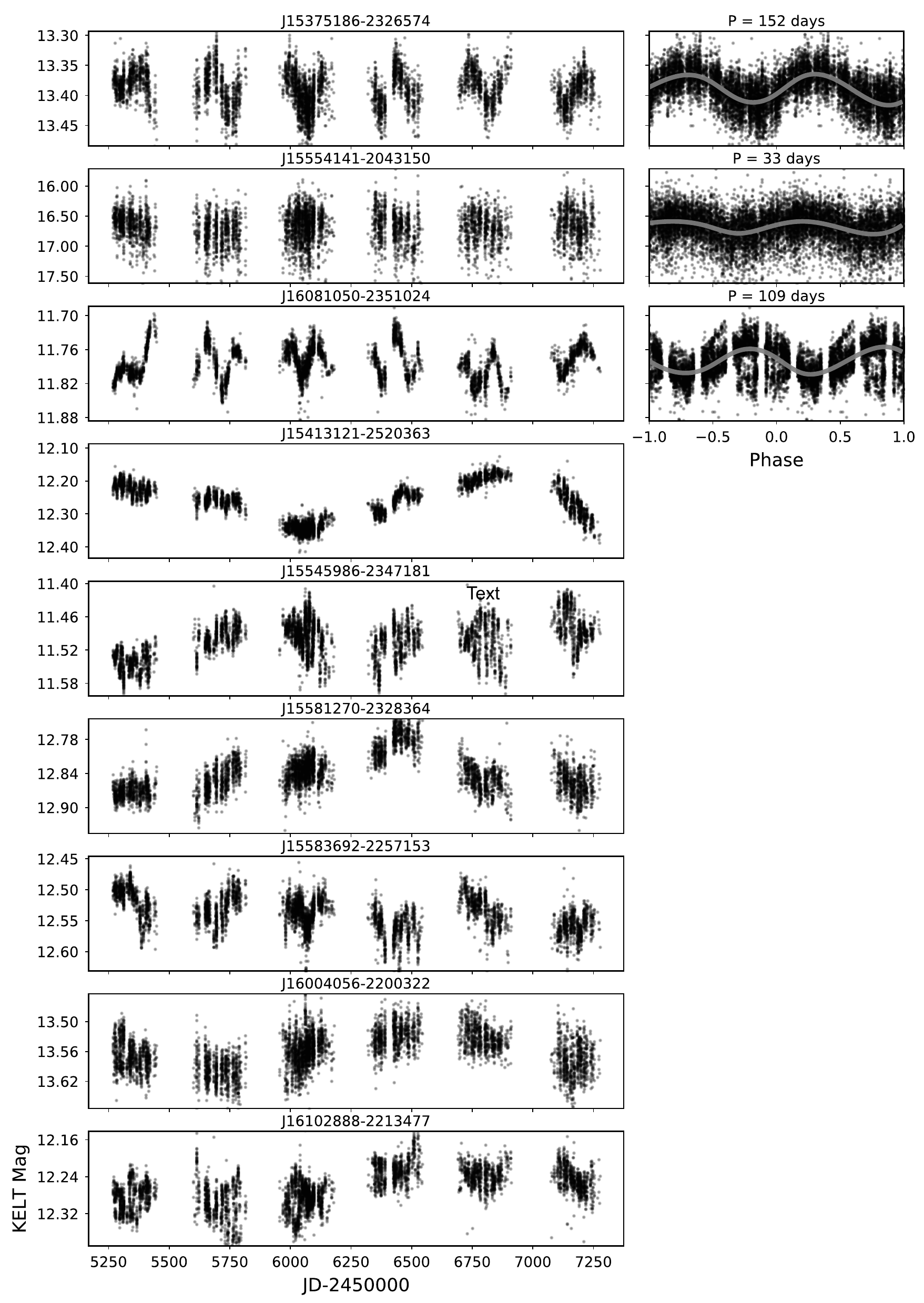}
\caption{The full KELT light curves (left) for the long-term variables discussed in Section~\ref{sec:LT}; source 2MASS names are given at the top of each panel. For those with identifiable periodic signals, we also show their phased light curves (right; replicated over two full phases) with smoothed medians (gray lines); the derived rotation periods are given at the top of each panel .}
\label{fig:LT}
\end{centering}
\end{figure*}
}

\subsection{Long-term Periodicity and Dimming}
\label{sec:LT}

Three sources in our sample (J15375186-2326574, J15554141-2043150, J16081050-2351024) exhibit periodic signals in their KELT light curves at $>20$~days, which is too long to be associated with stellar rotation for young ($\sim$10~Myr) pre-main sequence (PMS) stars. Six additional sources (J15413121-2520363, J15545986-2347181, J15581270-2328364, J15583692-2257153, J16004056-2200322, J16102888-2213477) show very long-term variability at $\gg100$~days, which may potentially be cyclic, although longer time baselines are needed to confirm this. These nine sources are discussed below and their light curves are shown in Figure~\ref{fig:LT}. We consider the possibility of their long-term variability being related to stellar activity cycles in Section~\ref{sec:activity}.

{\bf J16081050-2351024} has an F3 spectral type and hosts a ``debris/evolved transitional" disk \citep{Luhman:2012}. Its KELT light curve shows a 109-day period and \cite{2015Ripepi} have used its {\it K2}/C2 light curve to classify the star as a $\gamma$~Dor variable exhibiting multiple low-amplitude (ppt), high-frequency ($\lesssim1$~d$^{-1}$) pulsations.

{\bf J15554141-2043150} was identified as an Upper Sco member of M1 spectral type by \cite{Rizzuto:2015}, who found clear lithium absorption ($EW({\rm Li})=0.64$~\AA) and strong H$\alpha$ emission ($EW({\rm H\alpha})=-6.1$~\AA) consistent with a PMS star, though no WISE infrared excess indicative of a disk. The source exhibits a 33-day period in the KELT data and a clear 1-day period in its {\it K2}/C2 light curve with 10\% amplitude. 

{\bf J15375186-2326574} was identified as a young star of K2V spectral type by \cite{Preibisch:1998}; however, due to its weak lithium absorption ($EW({\rm Li})=0.12$~\AA), they classified it as a candidate young cluster member rather than a PMS star. This source exhibits a 153-day period in our KELT data and will be observed during {\it K2}/C15. No shorter periods are found in the KELT data.

The six sources showing very long-term variability are all classified as G-type stars. J15413121-2520363 and J16102888-2213477 are classified as active G-type sub-giants \cite[G9IVe and G7IVe, respectively;][]{2006Torres}, though their apparent low surface gravity is likely rather due to their PMS status. Indeed, both the KELT and {\it K2}/C2 data of J16102888-2213477 contain a 2.29-day periodic signal, which is presumably due to stellar rotation. While no rotation period is found in the KELT data for J15413121-2520363, the star will be observed during {\it K2}/C15. J15545986-2347181, J16004056-2200322, and J15581270-2328364 all have G dwarf spectral types \cite[G3, G9, and G6, respectively;][]{2006Torres} with stellar rotation periods found with KELT and {\it K2}/C2 data of 1.06, 2.71, and 1.72~days, respectively.  These sources will all be re-observed in {\it K2}/C15. Finally, J15583692-2257153 is a G7 star  \citep{Luhman:2012} hosting a massive disk with a centrally depleted cavity that has been resolved at sub-mm wavelengths by ALMA \citep{Barenfeld2016}.

\section{Discussion}
\label{sec:discussion}

\subsection{Comparing KELT and {\it K2} Periods}
\label{sec:k2}

\begin{figure}
\begin{centering}
\includegraphics[width=8.5cm]{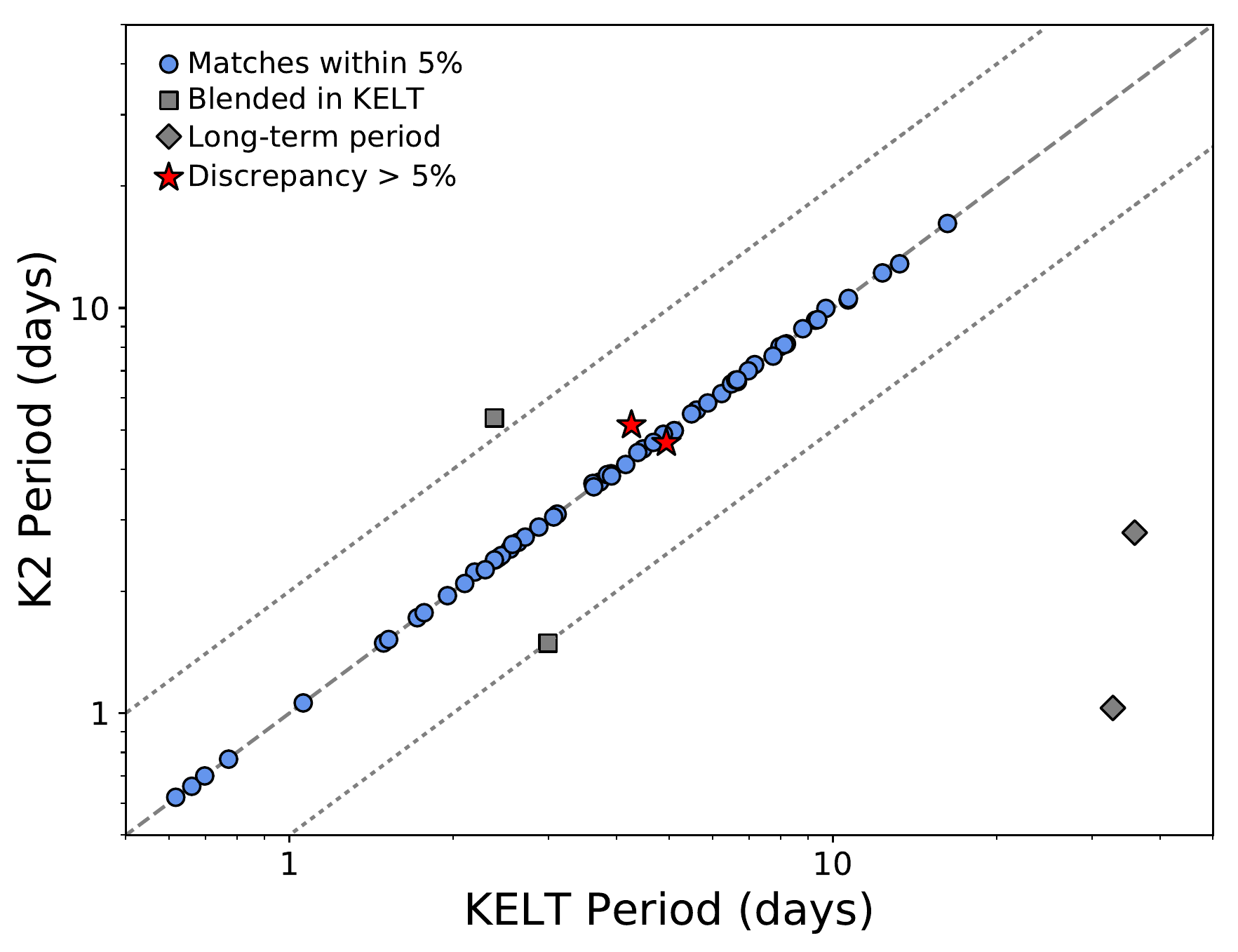}
\caption{Comparison of the rotational periods identified from KELT and {\it K2}/C2 data for the 67 sources in our sample for which we could derive significant periodic signals in both datasets (see Section~\ref{sec:k2}). The dashed line shows 1:1 agreement, while the dotted lines show the 2$\times$ and 0.5$\times$ discrepancies. Only 6 out of the 67 sources are discrepant at $>5$\% (see Section~\ref{sec:k2} for potential explanations of these discrepant sources).}
\label{fig:k2}
\end{centering}
\end{figure}

In order to test whether the periodic signals found in the KELT data (Section~\ref{sec:period}) are indeed stellar rotation periods, we can compare our derived values to the periodic signals found in the {\it K2}/C2 data, as any true stellar rotation periods should be retrieved in both datasets. To derive periodic signals from the {\it K2}/C2 data, we apply the LSP to the light curves reduced with the routines of \citet{Vanderburg:2014}, which are made publicly available on the Mikulski Archive for Space Telescopes (MAST).\footnote{\url{https://archive.stsci.edu/k2/hlsp/k2sff/search.php}} We do not apply any rejection criteria for avoiding diurnal or lunar signals (see Section~\ref{sec:period}), as these aliases should not occur in the space-based {\it K2} data. Figure~\ref{fig:k2} shows the comparison for the 67 targets in our sample for which we could derive periods from both KELT and {\it K2}/C2 datasets. The periods derived from KELT data match extremely well to those derived from {\it K2}/C2 data: only six sources are discrepant at $>5$\% and the median difference is just 0.4\%. Thus we can be confident that the periodic signals derived from the KELT data are indeed the stellar rotation periods in the vast majority of cases.

Two of the discrepant sources (J15554141-2043150, J16141107-2305362) have long periods that can be identified in the long-baseline KELT data (Section~\ref{sec:exotic}) but not in the short-baseline {\it K2}/C2 data. Two other discrepant sources are potential blends: J16082324-1930009 appears to have the period of a brighter nearby star, while J15573430-2321123 shows the 2$\times$ harmonic of its {\it K2}/C2 period. The two remaining sources (J16015149-2445249, J16125533-2319456) show only 6\% and 20\% discrepancies, respectively. J16015149-2445249 has a clear periodic signal in its {\it K2}/C2 data, but is relatively faint ($V=12.5$) for KELT, thus the discrepancy may be due to large uncertainty in the derived KELT period. J16125533-2319456 shows a secondary peak in its {\it K2}/C2 periodogram that is within 10\% of the derived KELT period, thus the discrepancy may be due to an unresolved binary. 

The excellent match between the periodic signals derived from KELT and {\it K2}/C2 data also demonstrates that long-term ground-based surveys like KELT are valuable tools for accurately deriving the rotation periods of young stars, so long as the target brightness is within the magnitude range for which the telescope is designed. Indeed, there are only 20 sources in our sample for which periods could be found with {\it K2}/C2 but not KELT; these are typically too faint for precise photometry with KELT (median $V=13.7$) or have very low-amplitude variability ($\ll$1\%).

\subsection{Rotational Evolution of PMS Stars}
\label{sec:mstar}

\begin{figure*}
\begin{centering}
\includegraphics[width=16.5cm]{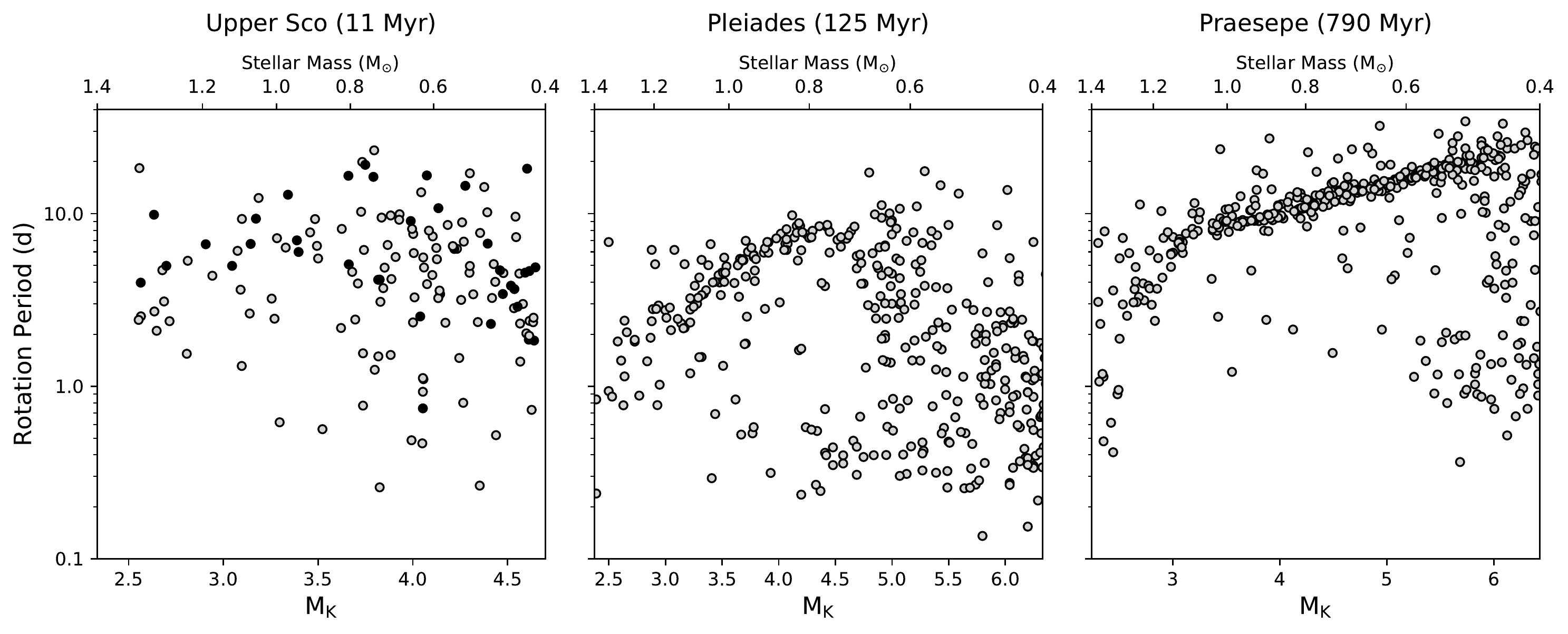}
\caption{Rotation period as a function of $M_{\rm K}$ (a proxy for stellar mass) for members Upper Sco (this work), the Pleiades \citep{2016Rebull}, and Praesepe \citep{2017Rebull}. $M_{\rm K}$ was derived from 2MASS photometry assuming distances of 145~pc for Upper Sco, 125~pc for Pleiades, and 180~pc for Praesepe. Approximate stellar masses are shown on the top axes and derived from \citet{2015Baraffe} isochrones for ages of 10~My for Upper Sco, 120~Myr for Pleiades, and 800~Myr for Praesepe. Upper Sco members likely hosting disks (as determined by excess infrared WISE emission) are indicated by black points.
}
\label{fig:mstar}
\end{centering}
\end{figure*}

Comparing the distributions of rotation periods for co-eval groups of young stars can inform early stellar evolution. In particular, such a comparison can provide insight into the spin-up of young stars as they contract toward the main sequence, the braking of stellar rotation via magnetized stellar winds that serve to remove angular momentum, and the exchange of angular momentum between the outer layers of stars with their interiors as well as any circumstellar disks \cite[e.g., see the recent review in][]{2015Gallet}. To this end, Figure~\ref{fig:mstar} shows rotation period as a function of absolute 2MASS $K_{\rm S}$ magnitude ($M_{\rm K}$), a proxy for stellar mass, for Upper Sco compared to two older stellar clusters, the Pleiades \cite[125~Myr;][]{1998Stauffer} and Preasepe \cite[790~Myr;][]{2015Brandt}. 

The rotation periods for Pleiades and Praesepe members are taken from \citet{2016Rebull} and \citet{2017Rebull}, respectively, who derived these values from {\it K2} light curves. For Upper Sco, we estimate rotation periods for 208 candidate members observed during {\it K2}/C2 by again taking the publicly available light curves reduced using the routines of \citet{Vanderburg:2014} (see Section~\ref{sec:k2}), then identifying significant periods between 0.2 and 30~days using both the LSP and the auto-correlation function \citep[ACF;][]{McQuillan:2013}, with the ACF periods given priority. We also take from Table \ref{tab:CatalogueInfo} the KELT rotation periods for {\it K2}/C15 targets that have not yet been observed by {\it K2}. To calculate $M_K$ we assume distances of 145~pc for Upper Sco, 135~pc for the Pleiades \citep{2014Melis}, and 180~pc for Praesepe \citep{2009Leeuwen}.  We also show in Figure~\ref{fig:mstar} approximate stellar masses, which we derive by comparing $M_{\rm K}$ to \citet{2015Baraffe} stellar evolution models at the given cluster ages; the $x$-axis of each plot is scaled to cover the same inferred stellar mass range for all three clusters. The Upper Sco members exhibiting significant excess WISE emission and thus likely hosting circumstellar disks \citep{Luhman:2012} are marked as black points in Figure~\ref{fig:mstar}.

The wide spread in rotation periods seen for Upper Sco members is representative of a classic problem in early stellar evolution studies \citep[e.g.,][]{Krishnamurthi:1997}. As young stars contract toward the main sequence, conservation of angular momentum dictates that spin rates increase. However, other mechanisms also influence stellar rotation rates, in particular disk-star ``locking" via magnetic torquing as well as ``braking" via magnetized stellar winds \citep[e.g.,][]{Landin:2016}. These competing mechanisms, in addition to differences in initial angular momenta, likely result in the large dispersion in rotation periods seen at the young ($\approx5$--11~Myr) age of Upper Sco.  However, Figure~\ref{fig:mstar} also shows the well-known evolution from this broad distribution of rotation periods to a tightly converged sequence that is clearly apparent by the age of the Pleiades (125~Myr) for $M_{\star}\gtrsim0.7~M_{\odot}$, and extends down to low-mass stars ($M_{\star}\gtrsim0.4~M_{\odot}$) by the age of Praesepe (790~Myr).

Given that most cool Upper Sco members still have faster rotation periods compared to stars along the loci seen in Praesepe, angular momentum loss via magnetized winds must be important at ages beyond that of Upper Sco; disk-star locking is unlikely to play a major role in spin-down, as only about 7\% of Upper Sco members are currently accreting \citep{2016Pecaut}, meaning that any disk-locking is \emph{fait accompli} (barring star formation within the past few Myr). Figure~\ref{fig:mstar} also suggests that even at the young age of Upper Sco, stars more massive than $\approx0.8~M_{\odot}$ may be beginning to converge toward the loci seen at older ages. Lower-mass stars that are still evolving toward the main sequence will resist spin-down due to their more massive convective envelopes and higher moments of inertia (when ``unlocked" from the disk, and assuming that the inner radiative core and outer convective envelope of a star are only weakly coupled). This could explain the progressive disappearance of the dispersion of rotation periods as a function of stellar mass seen in Figure~\ref{fig:mstar}. Reality may be more complex, however, since the h~Persei region at 13~Myr shows no decline in the dispersion of rotation period with stellar mass \citep[see Figure 11 in][]{2013Moraux}. 

We also note that the distribution of rotation periods for Upper Sco members has an envelope with an upper edge of $\approx 10-20$~days for low-mass ($\lesssim0.8M_{\odot}$) stars, decreasing to shorter periods for higher-mass stars. A similar distribution was found for Upper Sco in the ground-based SuperWASP data analyzed in \citet{Mellon:2017}, although with an upper bound of 8~days (see also \citealt{Scholz:2015} for an extension to the brown dwarf regime at $M_K > 8$). The longer-period signals seen here could be produced by binaries as well as older interloping stars, although many of the longer-period stars in Upper Sco appear to host disks, suggesting that they are indeed young ($\lesssim10$~Myr).

\subsection{Stellar Activity Cycles}
\label{sec:activity}

Ten sources in our sample showed long-term ($>20$-day) variability in their KELT light curves (see Sections~\ref{sec:dippers} \& \ref{sec:LT}). Four of these sources exhibited clear periodic signals, which are too long to be explained by stellar rotation for $\sim$10~Myr old PMS stars. The other six sources showed hints of much longer-term (100s of days) cyclic behavior, which may become clearer with longer-baseline observations. 

Such long-term variability may be related to solar-like magnetic activity cycles that drive significant changes in spot coverage over time, which in turn modulate the brightness variations observed by KELT. Indeed, these long-term variables are strongly biased toward G-type and early K-type sources compared to the rest of our sample. Moreover, only two of these sources (J15583692-2257153, J16141107-2305362) host full disks, thus the long-term variability is unlikely to be solely due to cyclic changes in the interaction between the inner disk and the stellar magnetic field, which has been previously suggested to explain long-term periods seen in WASP data of T~Tauri stars \citep{2017Rigon}.

\section{Summary}
\label{sec:sum}

We presented and analyzed light curves spanning 5.5~years from the ground-based KELT survey for 131 candidate members of Upper Sco that are also targets of the {\it K2} mission. The long-baseline KELT data published in this paper (available online as electronic tables) provide a legacy dataset that complements the much higher-precision but shorter-duration photometric observations from {\it K2}, thereby enabling more complete studies of young star variability.

In our analysis, we found periodic signals for 89 stars in our sample, the vast majority of which are consistent with the rotation of spots on the surface of young stars. Namely, the periods span $\sim$0.5--10 days and have amplitudes of a few percent. We also identified stars exhibiting interesting variability apart from stellar rotation. In particular, we identified four sources exhibiting dipper/UXOR behavior, which is thought to be due to occultations of stars by dusty circumstellar material. The KELT data were able to identify new dippers and, when combined with {\it K2}/C2 data, also confirm long-term dipper behavior as well as the disappearance of dipper-like variability. Ten sources also showed potentially long-term cyclic behavior, which we interpreted as possibly being related to solar-like cycles of magnetic activity.

We also compared the periodic signals derived from KELT data to those derived from {\it K2}/C2 data for 67 sources. The excellent agreement confirmed that these periodic signals are likely stellar rotational signals, and also illustrated the power of long-term ground-based surveys for accurately deriving the rotation periods of young stars. Additionally, we compared the distribution of rotation period versus stellar mass in Upper Sco to those of the older Pleiades and Praesepe regions in order to analyze the angular momentum evolution of young stars from $\approx10$--800~Myr.
 
The continuation of long-term ground-based photometric monitoring surveys like KELT will be important for putting into context the high-precision data obtained from space-based missions beyond {\it K2}. In particular, the upcoming {\it Transiting Exoplanet Survey Satellite} ({\it TESS}; \citealp{Ricker:2014}) mission will obtain high-precision photometric observations of almost the entire sky for a duration of $\gtrsim27$ days. When {\it TESS} is in operation, KELT can provide $\gtrsim10$ years of complementary long-baseline data for studying the underlying processes driving young star variability and their relation to planet formation.

\section*{Acknowledgements}

Early work on KELT-North was supported by NASA Grant NNG04GO70G. J.E.R. was supported by the Harvard Future Faculty Leaders Postdoctoral fellowship. G.S. acknowledges the support of the Vanderbilt Office of the Provost through the Vanderbilt Initiative in Data-intensive Astrophysics (VIDA) fellowship. This work has made use of NASA's Astrophysics Data System and the SIMBAD database operated at CDS, Strasbourg, France. This document does not contain export controlled information and is under review for Unlimited External Release (URS268385). Part of the research was carried out at the Jet Propulsion Laboratory, California Institute of Technology, under a contract with the National Aeronautics and Space Administration.




\newpage
\bibliographystyle{mnras}
\bibliography{UpperSco_KELT}



\clearpage
\pagestyle{plain}
\onecolumn
\begin{landscape} 
\begin{ThreePartTable} 
\begin{TableNotes} 
\footnotesize 
\item $^{\dagger}$References: (1) = \cite{Luhman:2012}, (2) = \cite{Rizzuto:2015}, (3) = \cite{2016Pecaut} (4) = \cite{Preibisch:1998} 
\item $^{\ddagger}$Sources with periods that would have been rejected as potential aliases (see Section \ref{sec:period}) but were kept due to clear periodicity in the KELT data or confirmation of the period from {\it K2}/C2. 
\end{TableNotes} 
\renewcommand*{\arraystretch}{1.1} 
{\footnotesize 
\begin{longtable}{lccccrrrrrccrlr} 
\caption{Upper Sco Members Observed by KELT and {\it K2}} \\ 
\hline\hline 
2MASS ID & EPIC ID & Field & \multicolumn{1}{c}{RA$_{\rm J2000}$} & \multicolumn{1}{c}{Dec$_{\rm J2000}$} & \multicolumn{1}{c}{SpT} &\multicolumn{1}{c}{Disk} &\multicolumn{1}{c}{$V$} &\multicolumn{1}{c}{$K_{\rm S}$} &\multicolumn{1}{c}{RMS} &\multicolumn{1}{c}{Var} &\multicolumn{1}{c}{Blend} &\multicolumn{1}{c}{Period} &\multicolumn{1}{l}{Classification} &\multicolumn{1}{l}{Ref$^{\dagger}$} \\ 
\hline 
\endfirsthead 
\caption[]{continued} \\ 
\hline\hline 
\vspace{0.1mm} 
\endhead 
\hline \\ 
\insertTableNotes 
\endlastfoot 
J15195029-2313577 & 249386734 & 15 & 15:19:50.299 & -23:13:57.00 & K4 & 0 & 10.96 & 8.86 & 1.00 & 1 & 0 & 3.0611 & Periodic & (4) \\
J15221627-2652252 &  & 15 & 15:22:16.202 & -26:52:23.98 & K6 & 0 & 11.86 & 8.66 & 1.00 & 1 & 0 & 6.8304 & Periodic & (4) \\
J15313976-2520440 & 249225380 & 15 & 15:31:39.698 & -25:20:44.01 & G9 & 0 & 13.21 & 11.16 & 0.00 & 0 & 0 &  &  & (4) \\
J15331002-2323185 & 249374906 & 15 & 15:33:10.000 & -23:23:17.98 & G9 & 0 & 12.49 & 10.98 & 0.00 & 0 & 0 &  &  & (4) \\
J15344110-2322161 & 249376234 & 15 & 15:34:41.100 & -23:22:15.99 & G5 & 0 & 11.48 & 10.33 & 0.00 & 0 & 0 &  &  & (4) \\
J15355780-2324046 & 249373956 & 15 & 15:35:57.796 & -23:24:03.99 & K3 & 0 & 12.21 & 9.43 & 0.00 & 0 & 0 & 0.8030 & Periodic & (1) \\
J15375186-2326574 & 249370343 & 15 & 15:37:51.801 & -23:26:57.98 & K2 & 0 & 11.04 & 7.56 & 1.00 & 1 & 0 & 152.2776 & Periodic/Long-term & (4) \\
J15381253-2326096 & 249371327 & 15 & 15:38:12.499 & -23:26:09.99 & K3 & 0 & 10.76 & 8.98 & 0.00 & 0 & 0 &  &  & (4) \\
J15390696-2646320 & 249112324 & 15 & 15:39:06.957 & -26:46:32.08 & M0.5 & 0 & 12.78 & 8.67 & 0.00 & 0 & 0 & 4.8531 & Periodic & (2) \\
J15413121-2520363 & 249225568 & 15 & 15:41:31.212 & -25:20:36.31 & G9IVe & 0 & 9.95 & 7.24 & 1.00 & 1 & 0 &  & Long-term & (1) \\
J15424991-2536406$^{\ddagger}$ & 249205008 & 15 & 15:42:49.898 & -25:36:38.98 & G5 & 0 & 10.79 & 8.18 & 1.00 & 1 & 0 & 0.3994 & Periodic & (3) \\
J15435905-2622516 & 249144433 & 15 & 15:43:59.064 & -26:22:51.60 & K9Ve & 0 & 14.07 & 9.83 & 0.00 & 0 & 0 & 3.7295 & Periodic & (3) \\
J15441334-2522590 & 249222494 & 15 & 15:44:13.300 & -25:22:58.00 & M1 & 0 & 13.47 & 9.08 & 0.00 & 0 & 0 & 6.4121 & Periodic & (1) \\
J15450970-2512430 & 249235696 & 15 & 15:45:09.708 & -25:12:42.98 & M1.5 & 0 & 13.52 & 9.72 & 0.00 & 0 & 0 & 2.5117 & Periodic & (2) \\
J15453913-2520449 & 249225361 & 15 & 15:45:39.100 & -25:20:44.98 & G3 & 0 & 11.43 & 9.13 & 0.00 & 0 & 0 & 1.8087 & Periodic & (4) \\
J15465590-2016022 & 249619259 & 15 & 15:46:55.898 & -20:16:01.99 & G0 & 0 & 11.40 & 10.16 & 0.00 & 0 & 0 &  &  & (4) \\
J15492100-2600062 & 249174457 & 15 & 15:49:21.000 & -26:00:06.30 & K0.5 & 0 & 11.00 & 7.91 & 1.00 & 1 & 0 &  &  & (1) \\
J15492289-2606068 & 249166617 & 15 & 15:49:22.800 & -26:06:06.98 & G1 & 0 & 12.04 & 10.08 & 0.00 & 0 & 0 &  &  & (4) \\
J15495979-2509033 & 249240256 & 15 & 15:49:59.786 & -25:09:03.56 & A2V & 1 & 9.25 & 7.89 & 0.00 & 0 & 0 &  &  & (1) \\
J15505641-2534189 & 203604834 & 15 & 15:50:56.419 & -25:34:18.98 & G0 & 0 & 9.75 & 7.91 & 0.00 & 0 & 0 & 0.9265 & Periodic & (1) \\
J15514535-2456513 & 203782856 & 15 & 15:51:45.360 & -24:56:51.36 & K3IV & 0 & 12.37 & 9.53 & 0.00 & 0 & 0 & 1.5284 & Periodic & (3) \\
J15523122-2633529 & 203307585 & 15 & 15:52:31.226 & -26:33:52.95 & M0 & 0 & 12.31 & 8.98 & 0.00 & 0 & 0 & 3.3018 & Periodic & (1) \\
J15545986-2347181 & 204069863 & 02/15 & 15:54:59.868 & -23:47:18.16 & G3V & 0 & 9.01 & 7.03 & 1.00 & 1 & 1 & 1.0601 & Periodic/Long-term & (1) \\
J15550213-2149434 & 204562176 & 02/15 & 15:55:02.140 & -21:49:43.50 & M0.0 & 0 & 12.25 & 8.64 & 0.00 & 0 & 1 & 4.1558 & Periodic & (2) \\
J15550624-2521102 & 203667805 & 02 & 15:55:06.242 & -25:21:10.22 & M1 & 0 & 12.57 & 8.51 & 0.00 & 0 & 0 & 3.9047 & Periodic & (1) \\
J15551704-2322165 & 204174563 & 02/15 & 15:55:17.040 & -23:22:16.60 & M2.5 & 1 & 13.85 & 9.33 & 0.00 & 0 & 1 &  &  & (1) \\
J15551758-2322036 & 204175508 & 02/15 & 15:55:17.584 & -23:22:03.75 & A4IV/V & 0 & 8.43 & 7.39 & 0.00 & 0 & 1 &  &  & (1) \\
J15552980-2544499 & 203553934 & 02 & 15:55:29.812 & -25:44:49.99 & M0.5 & 0 & 13.43 & 9.14 & 0.00 & 0 & 0 &  &  & (2) \\
J15554141-2043150 & 204819202 & 02 & 15:55:41.412 & -20:43:15.09 & M1.0 & 0 & 13.78 & 9.45 & 0.00 & 0 & 0 & 32.7250 & Periodic/Long-term & (2) \\
J15554883-2512240 & 203710077 & 02 & 15:55:48.835 & -25:12:24.08 & G3V & 1 & 10.50 & 8.29 & 1.00 & 1 & 0 & 3.7226 & Periodic/Dipper & (1) \\
J15562941-2348197 & 204065688 & 02/15 & 15:56:29.400 & -23:48:19.00 & M1.5 & 0 & 12.79 & 8.74 & 0.00 & 0 & 1 & 7.9855 & Periodic & (1) \\
J15563719-2332012 & 204133258 & 02/15 & 15:56:37.200 & -23:32:00.99 & G1 & 0 & 12.17 & 10.61 & 0.00 & 0 & 1 &  &  & (4) \\
J15564769-1950077 & 205010817 & 02 & 15:56:47.690 & -19:50:07.62 & M2.0 & 0 & 13.37 & 8.95 & 0.00 & 0 & 0 & 2.6345 & Periodic & (2) \\
J15565545-2258403 & 204274993 & 02/15 & 15:56:55.461 & -22:58:40.40 & M0 & 0 & 13.42 & 9.43 & 0.00 & 0 & 1 & 8.2100 & Periodic & (1) \\
J15570234-1950419 & 205008763 & 02 & 15:57:02.299 & -19:50:40.99 & K7 & 0 & 11.70 & 8.37 & 0.00 & 0 & 0 & 2.5458 & Periodic & (1) \\
J15571674-2529192$^{\ddagger}$ & 203628765 & 02 & 15:57:16.740 & -25:29:19.32 & M0 & 0 & 12.54 & 8.86 & 0.00 & 0 & 0 & 0.6612 &  & (1) \\
J15571998-2338499 & 204104882 & 02/15 & 15:57:19.987 & -23:38:49.99 & M0 & 0 & 12.65 & 8.88 & 0.00 & 0 & 1 & 6.2388 & Periodic & (1) \\
J15572575-2354220 & 204040060 & 02/15 & 15:57:25.759 & -23:54:21.99 & M0.5 & 0 & 13.50 & 9.09 & 0.00 & 0 & 1 & 7.1756 & Periodic & (1) \\
J15573430-2321123 & 204179058 & 02/15 & 15:57:34.310 & -23:21:12.31 & M1 & 1 & 13.12 & 8.99 & 0.00 & 0 & 1 & 2.9901 & Periodic & (2) \\
J15581270-2328364 & 204147776 & 02/15 & 15:58:12.698 & -23:28:36.40 & G6 & 1 & 10.37 & 8.02 & 1.00 & 1 & 1 & 1.7185 & Periodic/Long-term & (1) \\
J15582965-2441445 & 203850326 & 02 & 15:58:29.599 & -24:41:43.98 & G0 & 0 & 13.22 & 11.70 & 0.00 & 0 & 0 &  &  & (4) \\
J15583692-2257153 & 204281213 & 02/15 & 15:58:36.921 & -22:57:15.33 & G7 & 1 & 10.15 & 7.05 & 1.00 & 1 & 1 &  & Long-term & (1) \\
J15590208-1844142 & 205230905 & 02 & 15:59:02.001 & -18:44:13.99 & K6.5 & 0 & 11.94 & 8.11 & 0.00 & 0 & 0 & 1.7647 & Periodic & (1) \\
J15595270-2526292 & 203642381 & 02 & 15:59:52.689 & -25:26:29.18 & M0.5 & 0 & 13.47 & 9.72 & 0.00 & 0 & 0 & 4.4724 & Periodic & (2) \\
J16004056-2200322 & 204519031 & 02/15 & 16:00:40.560 & -22:00:32.18 & G9 & 0 & 10.88 & 8.44 & 1.00 & 1 & 1 & 2.7147 & Periodic/Long-term & (1) \\
J16010519-2227311 & 204406748 & 02/15 & 16:01:05.193 & -22:27:31.17 & M3 & 0 & 13.11 & 8.75 & 0.00 & 0 & 1 &  &  & (1) \\
J16012233-1937222 & 205055104 & 02 & 16:01:22.336 & -19:37:22.29 & M1.5 & 0 & 14.14 & 9.56 & 0.00 & 0 & 0 & 8.1341 & Periodic & (2) \\
J16012563-2240403 & 204350686 & 02/15 & 16:01:25.624 & -22:40:40.33 & K3 & 0 & 13.41 & 8.52 & 1.00 & 1 & 1 & 2.3918 & Periodic & (1) \\
J16014743-2049457 & 204794876 & 02 & 16:01:47.400 & -20:49:45.01 & M0 & 0 & 13.10 & 8.61 & 0.00 & 0 & 0 & 1.4900 & Periodic & (1) \\
J16015149-2445249 & 203834337 & 02 & 16:01:51.492 & -24:45:24.94 & K7 & 0 & 12.48 & 8.48 & 0.00 & 0 & 0 & 4.9359 & Periodic & (1) \\
J16015790-2100353 & 204754061 & 02 & 16:01:57.897 & -21:00:34.99 & G2 & 0 & 11.36 & 9.76 & 0.00 & 0 & 0 &  &  & (4) \\
J16015822-2008121 & 204947015 & 02 & 16:01:58.224 & -20:08:12.15 & G7 & 0 & 10.45 & 7.67 & 1.00 & 1 & 0 & 1.7692 & Periodic & (1) \\
J16020039-2221237 & 204432860 & 02/15 & 16:02:00.388 & -22:21:23.90 & M1 & 1 & 13.12 & 8.84 & 0.00 & 0 & 1 & 2.8761 & Periodic & (1) \\
J16020845-2254588 & 204290833 & 02/15 & 16:02:08.450 & -22:54:59.11 & M1 & 0 & 13.59 & 9.55 & 0.00 & 0 & 1 &  &  & (1) \\
J16021045-2241280 & 204347304 & 02/15 & 16:02:10.500 & -22:41:29.00 & K6 & 0 & 11.61 & 8.06 & 0.00 & 0 & 1 &  &  & (1) \\
J16023814-2541389 & 203569230 & 02 & 16:02:38.136 & -25:41:39.12 & K8IVe & 0 & 13.05 & 9.33 & 0.00 & 0 & 0 & 3.1083 & Periodic & (3) \\
J16023910-2542078 & 203566940 & 02 & 16:02:39.096 & -25:42:06.98 & K7 & 0 & 12.17 & 9.12 & 0.00 & 0 & 0 & 3.1083 & Periodic & (1) \\
J16025123-2401574 & 204008342 & 15 & 16:02:51.199 & -24:01:55.99 & K4 & 1 & 12.12 & 8.93 & 0.00 & 0 & 1 & 3.5112 & Periodic & (1) \\
J16025243-2402226 & 204006744 & 15 & 16:02:52.399 & -24:02:21.98 & K0 & 0 & 11.43 & 7.64 & 0.00 & 0 & 1 & 3.5112 & Periodic & (1) \\
J16025396-2022480$^{\ddagger}$ & 204894575 & 02 & 16:02:53.959 & -20:22:48.07 & K6 & 0 & 12.70 & 8.19 & 1.00 & 1 & 0 & 1.9531 & Periodic & (1) \\
J16033550-2245560 & 204328600 & 02/15 & 16:03:35.500 & -22:45:56.08 & K0 & 0 & 11.13 & 8.36 & 1.00 & 1 & 1 & 2.4253 & Periodic & (1) \\
J16033829-1854076 & 205199316 & 02 & 16:03:38.299 & -18:54:07.70 & M1.0 & 0 & 14.36 & 9.54 & 0.00 & 0 & 0 & 0.7728 & Periodic & (2) \\
J16034187-2005577 & 204954915 & 02 & 16:03:41.870 & -20:05:57.69 & M2 & 0 & 14.03 & 9.49 & 0.00 & 0 & 0 &  &  & (1) \\
J16034334-2015314 & 204920926 & 02 & 16:03:43.339 & -20:15:31.39 & M2 & 0 & 14.18 & 9.72 & 0.00 & 0 & 0 & 5.6130 & Periodic & (1) \\
J16034695-2245246$^{\ddagger}$ & 204330803 & 02/15 & 16:03:46.951 & -22:45:24.69 & M1.5 & 0 & 13.38 & 9.10 & 0.00 & 0 & 1 & 13.2620 & Periodic & (2) \\
J16035767-2031055 & 204864076 & 02 & 16:03:57.600 & -20:31:05.01 & K5 & 1 & 12.87 & 8.37 & 1.00 & 1 & 0 & 3.8443 & Periodic & (1) \\
J16041893-2430392 & 203895983 & 02 & 16:04:18.933 & -24:30:39.31 & M2.5 & 1 & 13.47 & 8.85 & 0.00 & 0 & 0 & 2.4552 & Periodic & (2) \\
J16042097-2130415 & 204637622 & 02 & 16:04:20.973 & -21:30:41.54 & M3.5 & 0 & 14.28 & 9.43 & 1.00 & 1 & 1 & 5.1043 & Periodic/Dipper & (1) \\
J16042165-2130284 & 204638512 & 02 & 16:04:21.700 & -21:30:28.00 & K2 & 1 & 12.26 & 8.51 & 1.00 & 1 & 1 & 5.1043 & Periodic/Dipper & (1) \\
J16062196-1928445 & 205084870 & 02 & 16:06:22.048 & -19:28:44.50 & M0.5 & 1 & 13.34 & 8.62 & 0.00 & 0 & 0 & 10.1389 & Periodic & (1) \\
J16064385-1908056 & 205154017 & 02 & 16:06:43.864 & -19:08:05.49 & K6 & 1 & 13.55 & 9.20 & 0.00 & 0 & 0 & 6.9845 & Periodic & (1) \\
J16070356-2036264 & 204844509 & 02 & 16:07:03.561 & -20:36:26.49 & M0 & 0 & 12.38 & 8.10 & 1.00 & 1 & 0 &  &  & (1) \\
J16070393-1911338 & 205142483 & 02 & 16:07:03.900 & -19:11:31.99 & M1 & 0 & 13.97 & 9.22 & 0.00 & 0 & 0 & 6.6376 & Periodic & (1) \\
J16071778-2203364 & 204506777 & 02 & 16:07:17.788 & -22:03:36.46 & F8V & 0 & 8.74 & 7.05 & 0.00 & 0 & 0 &  &  & (1) \\
J16074006-2148426 & 204566199 & 02 & 16:07:40.058 & -21:48:42.69 & M0.5 & 0 & 13.21 & 9.67 & 0.00 & 0 & 0 & 6.6721 & Periodic & (2) \\
J16074449-2036030 & 204845955 & 02 & 16:07:44.493 & -20:36:03.06 & M4 & 0 & 14.35 & 9.08 & 0.00 & 0 & 0 &  &  & (1) \\
J16080141-2027416 & 204876697 & 02 & 16:08:01.420 & -20:27:41.68 & K8 & 0 & 13.74 & 9.29 & 0.00 & 0 & 0 & 9.2839 & Periodic & (1) \\
J16081050-2351024 & 204054556 & 02 & 16:08:10.509 & -23:51:02.44 & F3V & 1 & 9.17 & 7.83 & 1.00 & 1 & 0 & 109.3195 & Periodic/Long-term & (1) \\
J16081081-1904479 & 205164832 & 02 & 16:08:10.824 & -19:04:48.00 & K3IVe & 0 & 12.29 & 8.47 & 1.00 & 1 & 0 & 3.0627 & Periodic & (3) \\
J16081474-1908327 & 205152548 & 02 & 16:08:14.700 & -19:08:31.99 & K2 & 0 & 11.73 & 8.43 & 1.00 & 1 & 0 & 10.6598 & Periodic & (1) \\
J16082234-1930052 & 205080360 & 02 & 16:08:22.344 & -19:30:05.22 & M1 & 0 & 13.86 & 9.06 & 0.00 & 0 & 1 & 2.3829 & Periodic & (1) \\
J16082324-1930009 & 205080616 & 02 & 16:08:23.244 & -19:30:00.93 & K9 & 1 & 14.03 & 9.47 & 0.00 & 0 & 1 & 2.3829 & Periodic & (1) \\
J16082511-2012245 & 204932100 & 02 & 16:08:25.108 & -20:12:24.58 & M1 & 0 & 14.19 & 9.87 & 0.00 & 0 & 0 &  &  & (1) \\
J16083436-1911563 & 205141287 & 02 & 16:08:34.298 & -19:11:54.99 & K5 & 0 & 12.17 & 7.79 & 1.00 & 1 & 0 & 7.7576 & Periodic & (3) \\
J16083514-2045296 & 204810792 & 02 & 16:08:35.140 & -20:45:29.62 & F3V & 0 & 8.36 & 6.68 & 0.00 & 0 & 0 &  &  & (1) \\
J16084340-2602168 & 203468205 & 02 & 16:08:43.298 & -26:02:16.00 & G8 & 0 & 10.28 & 7.91 & 1.00 & 1 & 0 & 2.1912 & Periodic & (1) \\
J16084366-2522367 & 203660895 & 02 & 16:08:43.668 & -25:22:36.62 & F4V & 0 & 8.81 & 7.25 & 0.00 & 0 & 0 & 3.9124 & Periodic & (1) \\
J16090075-1908526 & 205151387 & 02 & 16:09:00.729 & -19:08:52.58 & K9 & 1 & 13.79 & 9.15 & 0.00 & 0 & 0 & 9.7017 & Periodic & (1) \\
J16090844-2009277 & 204942552 & 02 & 16:09:08.445 & -20:09:27.79 & M4 & 0 & 14.10 & 9.52 & 0.00 & 0 & 0 &  &  & (1) \\
J16091373-2001045 & 204972242 & 02 & 16:09:13.696 & -20:01:04.00 & G4 & 0 & 11.42 & 9.54 & 0.00 & 0 & 0 &  &  & (4) \\
J16092918-1852536 & 205203376 & 02 & 16:09:29.184 & -18:52:53.76 & K4IVe & 0 & 12.73 & 8.38 & 0.00 & 0 & 0 &  &  & (3) \\
J16093035-2443379 & 203842174 & 02 & 16:09:30.357 & -24:43:37.92 & M3.5 & 0 & 14.02 & 9.50 & 0.00 & 0 & 0 &  &  & (1) \\
J16093164-2229224 & 204398857 & 02 & 16:09:31.660 & -22:29:22.41 & M2.0 & 1 & 13.23 & 9.15 & 0.00 & 0 & 0 &  &  & (2) \\
J16093969-2200466 & 204518032 & 02 & 16:09:39.700 & -22:00:46.62 & M0.5 & 0 & 13.47 & 9.30 & 0.00 & 0 & 0 & 6.5025 & Periodic & (2) \\
J16094644-1937361 & 205054287 & 02 & 16:09:46.440 & -19:37:36.08 & M1 & 0 & 14.21 & 9.63 & 0.00 & 0 & 0 &  &  & (1) \\
J16095441-1906551 & 205157836 & 02 & 16:09:54.420 & -19:06:55.00 & M1 & 1 & 14.39 & 9.60 & 0.00 & 0 & 0 & 16.2387 & Periodic & (1) \\
J16100501-2132318 & 204630363 & 02 & 16:10:05.016 & -21:32:31.88 & M0.0 & 1 & 13.04 & 8.95 & 0.00 & 0 & 0 & 6.6224 & Periodic & (2) \\
J16101729-1910263 & 205146246 & 02 & 16:10:17.304 & -19:10:26.40 & K5IVe & 0 & 12.72 & 8.53 & 0.00 & 0 & 0 &  &  & (3) \\
J16102653-2756293 & 202904813 & 02 & 16:10:26.544 & -27:56:29.40 & M3Ve & 0 & 13.93 & 9.57 & 0.00 & 0 & 0 &  &  & (3) \\
J16102857-1904469 & 205164892 & 02 & 16:10:28.576 & -19:04:46.99 & M3 & 1 & 13.69 & 8.71 & 0.00 & 0 & 0 & 6.6657 & Periodic & (1) \\
J16102888-2213477 & 204464828 & 02 & 16:10:28.879 & -22:13:47.78 & G7IVe & 0 & 9.71 & 7.23 & 1.00 & 1 & 0 & 2.2922 & Periodic/Long-term & (1) \\
J16103196-1913062 & 205137430 & 02 & 16:10:31.960 & -19:13:05.98 & K7 & 0 & 13.51 & 8.99 & 0.00 & 0 & 0 & 12.3305 & Periodic & (1) \\
J16110479-2333166 & 204127977 & 02 & 16:11:04.797 & -23:33:16.59 & M0.0 & 0 & 13.38 & 9.66 & 0.00 & 0 & 0 & 4.8760 & Periodic & (2) \\
J16110890-1904468 & 205164889 & 02 & 16:11:08.908 & -19:04:46.84 & K2 & 0 & 12.05 & 7.69 & 1.00 & 1 & 0 & 3.7946 & Periodic & (1) \\
J16112601-2631558 & 203318214 & 02 & 16:11:26.028 & -26:31:55.88 & M2.0 & 1 & 14.43 & 9.57 & 0.00 & 0 & 0 &  &  & (2) \\
J16114387-2526350 & 203641931 & 02 & 16:11:43.872 & -25:26:35.16 & K3IV(e) & 0 & 11.77 & 8.74 & 0.00 & 0 & 0 & 8.7973 & Periodic & (3) \\
J16115266-2232421 & 204384743 & 02 & 16:11:52.658 & -22:32:42.10 & A3V & 0 & 8.24 & 7.27 & 0.00 & 0 & 0 &  &  & (1) \\
J16115633-2304051 & 204251947 & 02 & 16:11:56.294 & -23:04:04.00 & M1 & 0 & 12.71 & 8.81 & 1.00 & 1 & 0 & 5.4915 & Periodic & (1) \\
J16115927-1906532 & 205157937 & 02 & 16:11:59.277 & -19:06:53.28 & K0 & 0 & 11.59 & 8.09 & 1.00 & 1 & 0 & 3.6157 & Periodic & (1) \\
J16120505-2043404 & 204817605 & 02 & 16:12:05.049 & -20:43:40.51 & M1.0 & 1 & 13.74 & 9.06 & 0.00 & 0 & 0 & 9.3735 & Periodic & (2) \\
J16123531-2034339 & 204851310 & 02 & 16:12:35.323 & -20:34:33.99 & M0.5 & 0 & 13.73 & 9.30 & 0.00 & 0 & 0 &  &  & (2) \\
J16123604-2723031 & 203063710 & 02 & 16:12:36.052 & -27:23:03.19 & K4.0 & 0 & 11.26 & 7.23 & 0.00 & 0 & 0 &  &  & (2) \\
J16124051-1859282 & 205182220 & 02 & 16:12:40.512 & -18:59:28.28 & K6 & 0 & 10.97 & 7.49 & 1.00 & 1 & 0 & 1.5224 & Periodic & (1) \\
J16124123-1949380 & 205012599 & 02 & 16:12:41.236 & -19:49:38.02 & M1 & 0 & 13.39 & 8.90 & 0.00 & 0 & 0 & 3.6293 & Periodic & (1) \\
J16124682-2213317 & 204465961 & 02 & 16:12:46.824 & -22:13:31.80 & K3IV(e) & 0 & 11.20 & 8.03 & 0.00 & 0 & 1 & 10.6760 & Periodic & (3) \\
J16125533-2319456 & 204185181 & 02 & 16:12:55.334 & -23:19:45.69 & G2V & 1 & 9.36 & 7.29 & 0.00 & 0 & 0 & 4.2582 & Periodic & (1) \\
J16130271-2257446 & 204279085 & 02 & 16:13:02.700 & -22:57:42.98 & K4.5 & 0 & 11.72 & 8.46 & 0.00 & 0 & 0 & 2.1012 & Periodic & (1) \\
J16131158-2229066 & 204399980 & 02 & 16:13:11.584 & -22:29:06.68 & A8III/IV & 1 & 8.87 & 6.69 & 1.00 & 1 & 0 &  & Dipper & (1) \\
J16131858-2212489 & 204468888 & 02 & 16:13:18.499 & -22:12:48.02 & K0 & 0 & 10.40 & 7.43 & 1.00 & 1 & 0 & 5.8812 & Periodic & (1) \\
J16133644-2326270 & 204156820 & 02 & 16:13:36.441 & -23:26:26.98 & M2.5 & 0 & 14.07 & 9.31 & 0.00 & 0 & 0 &  &  & (2) \\
J16134366-2214594 & 204459941 & 02 & 16:13:43.656 & -22:14:59.64 & K7IVe & 1 & 13.22 & 9.11 & 0.00 & 0 & 0 & 4.6731 & Periodic & (3) \\
J16134781-2747340 & 202947197 & 02 & 16:13:47.820 & -27:47:34.00 & F4.5 & 1 & 9.30 & 8.01 & 0.00 & 0 & 0 &  &  & (2) \\
J16140733-2217321 & 204449165 & 02 & 16:14:07.348 & -22:17:32.20 & M1.0 & 0 & 13.33 & 9.10 & 0.00 & 0 & 0 & 0.6178 & Periodic & (2) \\
J16141107-2305362 & 204245509 & 02 & 16:14:11.001 & -23:05:35.01 & K2 & 1 & 10.71 & 7.46 & 1.00 & 1 & 0 & 35.8945 & Periodic/Dipper & (1) \\
J16145918-2750230 & 202933888 & 02 & 16:14:59.203 & -27:50:22.02 & G8 & 1 & 10.98 & 8.69 & 1.00 & 1 & 0 & 4.3735 & Periodic & (1) \\
J16150927-2345348 & 204076987 & 02 & 16:15:09.280 & -23:45:35.02 & F3V & 1 & 9.49 & 8.03 & 0.00 & 0 & 0 &  &  & (1) \\
J16151948-2540119 & 203576256 & 02 & 16:15:19.488 & -25:40:12.00 & M1.5 & 0 & 14.30 & 9.12 & 0.00 & 0 & 0 & 0.6985 & Periodic & (2) \\
J16153215-2530310 & 203623026 & 02 & 16:15:31.996 & -25:30:29.98 & G2 & 0 & 10.73 & 8.24 & 0.00 & 0 & 0 & 2.5721 & Periodic & (4) \\
J16153311-2707587 & 203136724 & 02 & 16:15:33.110 & -27:07:58.80 & M0.5 & 0 & 14.02 & 9.76 & 0.00 & 0 & 0 & 3.2574 & Periodic & (2) \\
J16153587-2529008 & 203630298 & 02 & 16:15:35.856 & -25:29:00.99 & K5 & 0 & 12.89 & 8.74 & 0.00 & 0 & 0 & 1.8903 & Periodic & (1) \\
J16162191-2809504 & 202842502 & 02 & 16:16:21.919 & -28:09:50.47 & A1 & 0 & 9.13 & 7.93 & 0.00 & 0 & 0 &  &  & (1) \\
\label{tab:CatalogueInfo}
\end{longtable}} 
\end{ThreePartTable} 
\clearpage
\end{landscape}


\bsp	
\label{lastpage}
\end{document}